

\magnification=1200
\hsize=16 true cm
\vsize=22 true cm

\centerline{\bf ON THERMAL COMPTONIZATION IN e$^\pm$ PAIR PLASMAS}

\vskip 0.3 true cm
\centerline{Gabriele Ghisellini$^1$ and Francesco Haardt$^{2,3}$}
\vskip 0.3 true cm

\centerline{1: Osservatorio di Torino, Pino Torinese, 10025 Italy}

\centerline{E--mail: 32065::ghisellini; ghisellini@to.astro.it}

\centerline{2: ISAS/SISSA, Via Beirut 2--4, 34014 Trieste, Italy}

\centerline{E--mail: 38028::haardt; haardt@tsmi19.sissa.it}

\centerline{3: present address: STScI, 3700 San Martin Dr., Baltimore, MD
21218}

\centerline{E--mail: 6559::haardt; haardt@kepler.stsci.edu}

\vskip 0.4 true cm
\centerline{\bf ABSTRACT}
\vskip 0.3 truecm
\noindent
We study $e^{\pm}$ pair plasmas in pair equilibrium, which emit high
energy radiation by thermal Comptonization of soft photons.
We find that the maximum luminosity to size ratio of the source (i.e.
the compactness) depends not only on the hot plasma temperature, but
also on the spectral index of the resulting Comptonized spectrum.
In the observationally interesting range, sources of same compactness
can be hotter if their spectrum is steeper.
Instruments observing in the 50--500 keV energy range, such as OSSE
on board CGRO, and especially the future SAX satellite,
can be more successful in detecting sources moderately
steep, the flattest sources being characterized by an high energy
cut--off at too low frequencies.
For any given pair of values of spectral index and temperature,
Comptonization theory alone fixes the $ratio$ of the compactnesses in
hard and soft photons.
However, if the source is pair dominated, the absolute values of the two
compactnesses are fixed.
Therefore there is a a one--to--one correspondence between the physical
parameters of the source (the compactnesses in soft and hard photons)
and the observable quantities (spectral index and temperature).
This correspondence can be extremely useful in interpreting the physical
behaviour of the sources, especially during variations, and can help
discriminating between different models for the high energy emission of
compact sources.
\vskip 0.3 true cm
{\it Subject headings:}
radiation mechanisms: thermal --
galaxies: Seyfert --
gamma rays: theory --
plasmas --
X--rays: general
\vskip 0.3 truecm
{\centerline{\it to be published in Ap.J. Letters}}

\vfill
\eject
\centerline{\bf 1. INTRODUCTION}
\vskip 0.5 truecm

\noindent
The recent observations of OSSE on the Compton Gamma Ray Observatory
have shown a break or an exponential cut--off in the high energy X--ray
spectrum of Seyfert galaxies (Maisack et al. 1993, Cameron et al. 1993,
Madejski et al. in preparation).


These observations can be interpreted in the framework of thermal
Comptonization models, even if a non--thermal origin of the high energy
radiation cannot be ruled out (Zdziarski, 1994).
Thermal models have been recently successful in explaining the X--ray
background as the sum of the emission of Seyfert galaxies, even if it is not
yet clear which type of Seyfert galaxies contribute the most
(see e.g. Madau, Ghisellini \& Fabian 1993;
Zdziarski, Zycki \& Krolik 1993).

The fact that there exists a relatively narrow range of temperatures of
emitting plasmas in AGNs can be understood on the basis of the
thermostat effect of electron--positron pairs, after the classic papers
of Bisnovatyi-Kogan, Zel'dovich \& Sunyaev (1971), Svensson (1982, 1983,
1984), Lightman (1982) and Zdziarski (1985).

The main result of these studies is that, for any given luminosity
to size ratio of the source, there exists a maximum temperature at
which the emitting plasma can be in steady state.
If the temperature is greater than this
maximum limit, photon--photon, photon--particle and particle--particle
collisions
produce pairs that cannot annihilate at the same rate.
For the same reason, increasing the luminosity to size ratio
(the compactness), the temperature is bound to decrease.

Furthermore there are indications that the high energy emission
of AGNs in general, and Seyfert galaxies in particular, comes from
the inner regions surrounding the black hole, and that the
luminosity is greater than few per cent of the Eddington limit
(Padovani 1989, Done \& Fabian 1990).
We therefore believe that AGNs are characterized by a relatively narrow
range of compactnesses and, hence, of temperatures.

However, there can be differences among different AGNs, and, most
notably, there are differences in the spectral indices of their X--ray
emission.
Although the 2--10 keV spectral indices $\alpha_x$ of Seyfert 1 galaxies
and quasars have a small scatter around the mean value, $\alpha_x$
ranges from 0.4 to values steeper than 1 (Turner et al. 1990,
Comastri et al. 1992).

In this paper we investigate the behaviour of sources in pair
equilibrium characterized by different spectral indices by assuming
thermal Comptonization of soft photons as the main radiation mechanism.
In the framework of these models different spectral indices are due to
different ratios $L_s/L_h$ between the luminosity in soft
photons and the luminosity produced by the hot particles.
We therefore investigate the equilibrium states of hot plasmas characterized by
different compactnesses and different $L_s/L_h$ ratios.

The main results of this work are: a relation between the spectral index
$\alpha_x$ and the temperature $T$ of plasmas in pair equilibrium, and
a one--to--one correspondence between $\alpha$, $T$ and the
compactnesses in hard and soft photons.

These results can have important consequences:

i) for predicting a correlation between spectral index and cut--off energy in
pair dominated AGNs, in the sense already observed for NGC 4151 (Jourdain et
al. 1992 and Maisack et al. 1993,
Zdziarski, Lightman \&
Maciolek--Nied\'zwiecki, 1993) and IC 4329A (Madejski et al. 1994);

ii) for predicting the high energy emission, in the OSSE range, of
sources with different spectral indices in the 2--10 keV band;

iii) for interpreting the behaviour of the overall spectrum of
individual sources during spectral and flux variations.

\vskip 1 true cm
\centerline{\bf 2. THERMAL COMPTONIZATION AND PAIR PRODUCTION}
\vskip 0.5 true cm

\noindent
Thermal Comptonization has been extensively studied in the past (see
e.g. Sunyaev \& Titarchuk 1980, Pozdnyakov, Sobol' \& Sunyaev 1983).

We assume that soft photons are homogeneously distributed throughout a
sphere of radius $R$, with a diluted black body spectral distribution of
dimensionless temperature $\Theta_{BB}\equiv kT_{BB}/(m_ec^2)$.
The sphere is homogeneously filled with hot plasma, with dimensionless
temperature $\Theta$, producing, via Comptonization, a luminosity $L_h$
corresponding to a compactness $\ell_h\equiv L_h \sigma_T/(Rmc^3)$ where
$\sigma_T$ is the Thomson cross section.
Analogously, $\ell_s$ is the compactness of the soft photon source.

The scattering optical depth, $\tau_T$, and the temperature, $\Theta$,
of the hot plasma uniquely determine the Comptonized photon energy
distribution, characterized by a power law and a cut--off at
$h\nu/(m_ec^2)\sim \Theta$.
Therefore we can treat the temperature $\Theta$ and the spectral index
$\alpha$ (instead of $\Theta$ and $\tau_T$) as the two variables.

Although sufficient to completely determine the shape of the Comptonized
photons, $\alpha$ and $\Theta$ do not yield its normalization, but only
the amplification factor, i.e. the ratio $\ell_h/\ell_s$.
Symbolically:
$$
(\alpha ,\, \Theta)\, \to \, \ell_h/\ell_s
\eqno(1)
$$
Note that the reverse is not true, since the same amplification can be
achieved with a range of values of $\alpha$--$\Theta$.

In order to determine the absolute value of $\ell_h$ and $\ell_s$
one has to consider the effects of $e^{\pm}$ pair production.

The main result of hot plasma studies is that the compactness of a
source in equilibrium at a given temperature cannot reach arbitrary
large values.
In fact, as long as the compactness is very small, an increase in the
heating rate corresponds to an increased mean energy per particle, and
therefore to an increased temperature.
But when the temperature starts to be relativistic, pair production is
important, and the number of particles in the thermal bath increases.
Since the available energy is now shared among more particles, the
temperature in this regime decreases as the heating rate (the compactness)
increases.
The equilibrium and steady state corresponds to pair balance: pairs are
destroyed at the same rate at which they are created.
For any given temperature, there is a $maximum$ compactness allowing pair
equilibrium, and for any given compactness, there is a $maximum$ allowed
temperature.
The precise value of the function $\ell_{h,max}(\Theta)$ depends on the
detail of the emission mechanism (bremsstrahlung, Comptonization,
cyclosynchrotron, and so on).
Hereafter we use the symbol $\ell_h$ to indicate the
maximum allowed value $\ell_h$.

If the main radiation mechanism is thermal Comptonization, any pair of values
of
$\alpha$ and $\Theta$ determines the maximum allowed compactness
$\ell_h$.
Symbolically we have:
$$
(\alpha ,\, \Theta)\, \iff \, (\ell_h,\, \ell_s)
\eqno(2)
$$
Note that now there is a one--to--one relation between the physical
parameters of the source ($\ell_h$ and $\ell_s$) and the observable
quantities ($\alpha$ and $\Theta$).

Following Zdziarski (1985, hereafter Z85) we calculated $\ell_h(\Theta)$
for different values of $\alpha$,
limiting ourselves to temperatures $\Theta<10$ ($kT<5$ MeV),
where particle--particle and particle--photon pair production
processes are less important than photon--photon interactions,
and can therefore be neglected.
The formulae used are described in Z85; the Comptonized spectrum is
described as the sum of a power law spectrum with an exponential cut--off
at $\Theta$ and a Wien spectrum at $\Theta$: $\ell_h = \ell_{pl} +\ell_W$.
The formulae of Z85 well describe the Comptonized spectrum
at high energies, but give a poor representation of the spectrum
at low energies.
To derive $\ell_h$ for $\alpha>1$ and $\ell_s$ (to be derived by photon
number conservation), a more detailed description of the Comptonized
spectrum at low energies is required.

We therefore computed the thermally Comptonized spectra in spherical geometry
by means of the full relativistic kernel (Jones 1968, Coppi \& Blandford 1990).
This gives the exact value $\ell_h/\ell_s$ for any $(\tau_T,\Theta)$.
We then computed a lower integration limit, $x_1$, for equation (1) in Z85
so that the integrated photon spectrum and luminosity coincide with the
exact values.
We checked that, with the given $\tau_T$ and
$\Theta$, the spectral indices used following the prescription of Z85
were correct.
We used $kT_{BB}=10$ eV and, neglected, for simplicity, pair escape.
Note that in the absence of pair escape there is an absolute maximum in
the allowed temperature ($\Theta_{max}=24$) derived considering
particle--particle interactions.

Our calculations are presented in Fig. 1.
As can be seen the curves $\ell_h(\Theta)$ have a sort of pivot
at $\Theta\sim 1$, being flatter for flatter spectral indices.

To qualitatively understand this behaviour, let us consider first a fixed
value of $\Theta$ in the region corresponding to $\ell_h>1$.
As $\Theta<1$, the only photons effective in pair production are in the
Wien tail of the spectrum (see e.g. fig. 4a of Z85).

Furthermore, if $\tau_T>1$, simple radiative transfer assures that the
compactness of the Wien component is fixed (it depends only on
$\Theta$), independently of $\tau_T$.
Increasing the total compactness then means to increase the compactness
of the power law component.
But the two compactnesses are related by the value of $\tau_T$:
their ratio $\ell_{pl}/\ell_{W}$ decreases as $\tau_T$ increases
(more photons are driven to the Wien peak).
Only a lower $\tau_T$ (i.e. steeper $\alpha$) therefore allows a greater
compactness of the power law component.

Consider now the region corresponding to $\ell_h<1$.
In this case both the total luminosity and the pair production process
are determined by $\ell_{pl}$, $\ell_W$ playing no role.
For fixed $\Theta$, $\ell_h$ can increase if:

i) the new equilibrium state yields a greater $\tau_T$ and therefore a
flatter spectral index.
In this case the luminosity is increased at high energies increasing the
pair production rate, which in turn yields the required increased
$\tau_T$.

ii) the new equilibrium state yields a much lower value of $\tau_T$
and a much steeper spectrum.
In this case the luminosity is increased at low energies and decreased
at high energies, the pair production rate is decreased,
and the corresponding $\tau_T$ decreases.

The first solution gives $\alpha<1$ while the second one yields
$\alpha>1$.
Although both these solutions are consistent with pair balance and
Comptonization, the latter one exists only in a range of $\Theta$. This
range becomes larger for small values of the minimum energy of the
Comptonized spectrum $x_1$ (where most of the luminosity is for
$\alpha>1$).
For the adopted $\Theta_{BB}$, the two solutions exist only in the small
range $1<\Theta<2$, as shown in Fig. 1.

Fig. 1 shows that in the interesting (from the observational point of
view) parameter range $0.1<\Theta<1$ pair dominated sources with same
compactness should show a clear correlation between the spectral index
and the cut--off energy: the flatter the spectral index, the lower the
temperature.

Note that, as shown in Fig. 1, the maximum possible compactnesses for
$kT \sim 50$ keV are extremely large, but these corresponds to completely
pair dominated sources: if the hot emitting plasma has also a `normal',
electron--proton component, the compactness can be smaller (or the
temperature can be smaller than shown in Fig. 1, for the same
compactness).

\vskip 1 true cm
\centerline{\bf 3. `MAPPING'}
\vskip 0.5 true cm

\noindent
Thermal Comptonization together with pair plasma theory gives
a one--to--one relation between $\alpha$--$\Theta$ and
$\ell_s$--$\ell_h$.
This means that we can `map' the plane $\alpha$--$\Theta$ into the plane
$\ell_s$--$\ell_h$ (or equivalently $\ell_h/\ell_s$--$\ell_h$), and
viceversa.

This `mapping' links physical parameters, such as the soft and the hard
compactness, which completely characterize the source, with observable
quantities, such as the spectral index and cut--off energy.

Fig. 2 shows the results.
In the `starvness--compactness' plane (i.e $\ell_h/\ell_s$--$\ell_h$) we
have mapped the `temperature--spectral index' plane, plotting curves
for constant $\Theta$ and for constant $\alpha$.
We have restricted our analysis to the range which is observationally
interesting (i.e. $0.5<\alpha<1.5$ and $0.1<\Theta<1$ corresponding to
$51<kT<511$ keV).

It can be seen that $\ell_h/\ell_s$ is not exactly constant for a given
spectral index, but it has some weak dependence also on the temperature.
Moving along a curve with constant $\Theta$ in the direction of
steeper $\alpha$ we have that $\ell_h/\ell_s$ decreases
(as expected), while $\ell_h$ increases (as explained in the
previous section, and also shown in Fig. 1).

In this plane it is very easy to see what the equilibria states are
when the source varies.
For instance, suppose that a source initially in point labelled $A$ in Fig. 2
increases its power ($\ell_h$) by a factor 3.
The compactness in soft photons, $\ell_s$, can either remain constant
or vary together with $\ell_h$.
The latter case may indicate that there is some feedback between the
hot plasma producing the hard luminosity and the relatively cold plasma
producing the seed photons, as in a cold disk illuminated by a hot
corona.

The final states in the two cases are labelled $B$ and $C$ in Fig. 2.
As can be seen, point $B$ corresponds to a temperature lower by
a factor 1.5 and slightly steeper spectral index $\alpha$, while point $C$
corresponds to an even lower temperature (factor 2) and a flatter
$\alpha$.
All the possible intermediate cases $5<\ell_h/\ell_s<15$
are between points $B$ and $C$.

Another possibility is that the soft compactness varies even without any
change in $\ell_h$. This is possible only if reprocessing (as a source
of soft photons) is not important.
Point labelled $D$ corresponds to the final state of the source
initially in $A$, after an increase by a factor 3 of $\ell_s$.
The resulting equilibrium temperature is larger by a factor 1.2 and the
spectrum has steepened.


In conclusion, knowing how the spectral index and the temperature change,
one can know if $\ell_s$ is bound to follow the variation of $\ell_h$.
This is extremely important, as we can test models in which an important
role is played by reprocessing of the high energy radiation by cold matter
located near the hot gas.

\vskip 1 true cm
\centerline{\bf 4. DISCUSSION}
\vskip 0.5 true cm

\noindent
One of the main results of pair plasma studies is that the
maximum compactness allowed by pair equilibrium decreases as the
temperature increases.
This however refers to sources with {\it fixed} spectral index,
which roughly corresponds to fixed $\ell_h/\ell_s$.

We have shown that, from thermal Comptonization and pair plasma theories,
one can derive a one--to--one correspondence between the
observable quantities $\alpha$ and $kT$ and the physical
parameters $\ell_h$ and $\ell_s$.
The knowledge of any two of these 4 quantities univocally
determines the other two.

The spectral behaviour of a varying X--ray source depends on whether the
ratio $\ell_h/\ell_s$ or $\ell_s$ remains constant when $\ell_h$ varies.
If $\ell_h/\ell_s$ is constant, the spectral index slightly steepens for
increasing hard compactness and the temperature decreases.
If $\ell_s$ remains constant, the spectral index flattens for increasing
$\ell_h$ and the temperature decreases by a larger amount.

Observationally, there are indications that in some sources,
during large variations
of the X--ray flux, the X--ray spectral index remains
approximately constant (e.g. Nandra et al. 1991), or slightly
steepens (e.g. Yaqoob \& Warwick 1991).
This indicates that $\ell_h/\ell_s$ is nearly constant during variations,
strongly favoring models in which the soft component is dominated by
the reprocessed flux (see e.g. Haardt \& Maraschi 1991).
This in turn would imply the presence of cold matter [cloudlets
(Celotti, Fabian \& Rees 1992, Sviron \& Tsuruta 1993) or a cold
accretion disk (Lightman \& White 1988, George \&
Fabian 1991, Matt, Perola \& Piro, 1991)] very close to the illuminating
X--ray source.
The above conclusion does not depend strictly on the importance of
pairs, which instead can fix the temperature and the temperature change
during variations of the source.

Observations of the cut-off energy of the X--ray spectrum is of crucial
importance to establish if a source is pair dominated, and directly
yield an upper limit on the compactness as illustrated in Fig. 1.
In addition, if variations of the the flux corresponds to variations of
the high energy cut-off as illustrated in Fig. 2, we can derive the
compactness of the source, not only an upper limit, and therefore the
size of the emitting region.
Hence it is very important to coordinate observations in the 2--10 keV
band (which should yield the spectral index) and in the 50--500 keV band
(which should yield the value of the cut--off energy).
A good opportunity to pursue this program is presently offered by
ASCA and OSSE missions, and in future, by the SAX satellite.

If a class of sources, such as Seyfert galaxies, are pair dominated with
similar compactnesses, there should be a correlation between spectral
index and temperature.
In the range of compactness between 10 and 100, steeper spectra should
correspond to larger temperatures.
For $\ell_h=100$, the maximum temperature is 70, 200 and 400 keV for
spectral indices $\alpha=0.5$, 1 and 1.5, respectively.

At present, only two Seyfert galaxies have their spectral index and
cut--off energy reliably measured, i.e. NGC 4151 and IC 4329A, and it is
encouraging that the differences in their $\alpha_x$ and $kT$ are in the
sense discussed in this paper.
NGC 4151 has $\alpha=0.5$ and $kT\sim 50$ keV, while IC 4329A has
$\alpha\sim 1$ and $kT>15$0 keV.

The correlation discussed in this paper should be taken into account
when interpreting the X--ray background as due to Seyfert galaxies (and
quasars) of different spectral indices, since for each $\alpha$ a
different maximum temperature is indicated.

Finally, it is worth to point out that if a source has a flat $\alpha_x$
in the 2--10 keV band, it may have a cut--off at relatively small
energies, and therefore it may be invisible in the 100--200 keV band
(i.e. the OSSE band).
On the other hand, very steep sources ($\alpha_x>1$) have a very small
flux in the OSSE band, and are therefore difficult to detect even if the
cut--off energy is large.
As a consequence, the probability to detect sources at 100 keV should
peak for sources moderately steep ($\alpha_x\sim 1$) in the 2--10 keV
band.

\vskip 1 true cm
\centerline{\bf REFERENCES}
\vskip 0.5 true cm

\parindent=0 pt
\everypar={\hangindent=2.6pc}

Bisnovatyi-Kogan, G.S., Zel'dovich, Ya.B., \& Sunyaev, R.A.
1971, Astr. Zh., 48, 24 (Soviet Astr.. -- AJ, 15, 17)

Cameron, R.A. et al. 1993, Proceedings of the Compton symposium, ed.
N. Gehrels, St. Louis, in press

Celotti, A., Fabian, A.C., \& Rees, M.J. 1992, MNRAS, 255, 419

Comastri, A., Setti, G., Zamorani, G., Elvis, M., Giommi, P., Wilkes,
B.J., \& McDowell, J.C. 1992, ApJ, 384, 62

Coppi, P.S., \& Blandford, R.D. 1990, MNRAS, 245, 453

Done, C., \& Fabian, A.C. 1990, MNRAS, 240, 81

George, I.M., \& Fabian, A.C. 1991, MNRAS, 249, 352


Haardt, F., \& Maraschi, L. 1991, ApJ, 380, L51



Jones, F.C. 1968, Phys. Rev., 167, 1159

Jourdain, E., et al. 1992, 256, L38

Lightman, A.P. 1982, ApJ, 253, 842

Lightman, A.P., \& White, T.R.  1988, ApJ, 335, 738


Madau, P., Ghisellini, G., \& Fabian, A.C 1993, ApJ, 410, L10

Madejski, G., et al. 1994, ApJ, submitted

Maisack, M., et al. 1993, ApJ, 407, L61

Matt G., Perola G.C., Piro L., 1991, A\&A, 247, 25



Nandra, K., Pounds, K.A., Stewart, G.C., George, I.M., Hayashida, K.,
Makino, F., \& Ohashi, T. 1991, MNRAS, 248, 760.

Padovani, P. 1989, A \& A, 209, 27



Pozdnyakov, L.~A., Sobol', I.~M., \& Sunyaev, R.~A. 1983,
Ap. Space Sc. Rev., Vol. 2, p. 189

Sviron, R. \& Tsuruta, S. 1993, ApJ, 402, 420

Sunyaev R.A., \& Titarchuk, L.G. 1980, A\&A, 86, 121

Svensson, R. 1982, ApJ, 258, 335

Svensson, R. 1983, ApJ, 270, 300

Svensson, R. 1984, MNRAS, 209, 175




Turner, T.J., Weaver, K.A., Mushotzky, R.F., Holt, S.S., \& Madejski,
G.M. 1990, ApJ 381, 85

Yaqoob, T., \& Warwick, R.S. 1991, MNRAS, 248, 773

Zdziarski, A.A. 1985, ApJ, 289, 514 (Z85)

Zdziarski, A.A., Zycki, P.T., \& Krolik, J.H.,  1993, ApJ, 414, L81

Zdziarski, A.A., Lightman A.P., \&
Maciolek--Nied\'zwiecki, A. 1993, ApJ, 414, L93

Zdziarski, A.A. 1994, in proc. of the Second Compton Symp.,
in press

\vfill
\eject
\centerline{\bf FIGURE CAPTIONS}
\vskip 0.5 true cm

{\bf Fig. 1}: For any given temperature $\Theta\equiv kT/(m_ec^2)$
there exists a maximum possible compactness $\ell_h$ of a plasma
in pair equilibrium.
The exact shape of the function $\ell_h(\Theta)$ depends
on the spectral index resulting from the Comptonization process.
Flatter spectral indices correspond to flatter $\ell_h(\Theta)$.
Note that the curves (the labels indicate the value of the spectral index)
cross for $\ell_h\sim 1$.
For $\ell_h>1$, sources with the same compactness have
larger temperatures if their spectrum is steeper, and viceversa.

\vskip 0.5 true cm
{\bf Fig. 2}: In the plane $\ell_h/\ell_s$ vs $\ell_h$
(the starvness vs hard compactness plane) we have drawn the curves
for constant $\Theta$ (solid lines) and for constant $\alpha$
(dashed lines).
The temperature increases from right to left, $\alpha$ increases
(the spectrum steepens) from top to bottom, as labelled.
As illustration, consider a source, initially in point $A$, which
increases $\ell_h$ by a factor 3.
Its final equilibrium state will be between points $B$ and $C$,
if $\ell_s$ remains constant or increases by the same amount
of $\ell_h$.
If, instead, $\ell_s$ increases by a factor 3 without variations in
$\ell_h$, the final equilibrium state will correspond to the point
labelled $D$.

\bye